\documentclass[prl,twocolumn,floatfix,showpacs]{revtex4}
\usepackage{graphicx}
\usepackage{amssymb,amsfonts,amsmath}
\usepackage{color}

\newcommand{\bm}[1]{\mbox{\boldmath$#1$}}
\newcommand{\be}{\begin{equation}}
\newcommand{\ee}{\end{equation}}
\newcommand{\bea}{\begin{eqnarray}}
\newcommand{\eea}{\end{eqnarray}}
\newcommand{\non}{\nonumber}

\begin{document}
\title{Power-law relaxation behavior of an initially localized state in the spin-1/2 Heisenberg chain} 
\author{Tetsuo Deguchi, Pulak Ranjan Giri and Ryoko Hatakeyama$^\dagger$}

\affiliation{Department of Physics, Natural Science Division, Faculty of Core Research, 
Ochanomizu University, 2-1-1 Ohtsuka, Bunkyo-ku, Tokyo 112-8610, Japan}

\affiliation{$^\dagger$ Department of Physics, 
Graduate School of Science, 
University of Tokyo, \\ 
7-3-1 Hongo, Bunkyo-ku, Tokyo 113-0033, Japan} 

\begin{abstract} We present power-law relaxation behavior of the local magnetizations in the equilibration dynamics of the spin-1/2 Heisenberg spin chain as an isolated integrable quantum system. We perform the exact time evolution of the expectation values of the local spin operators by evaluating them with the determinant formula of the form factors.  
We construct such an initial quantum state that has a localized profile of the local magnetizations, and perform the exact time evolution over a very long period of time. We show that the local magnetization relaxes as some power of the time variable with no definite time scale, while the fidelity relaxes very fast with its relaxation time being proportional to the inverse of the energy width, i.e. the Boltzmann time. 
\end{abstract}
\date{\today}



\maketitle

\section{Introduction}

The non-equilibrium dynamics of isolated integrable quantum systems such as quantum quench has attracted much attention, recently \cite{McCoy1970,Rigol-GGE,Rigol-generic,SeiSuzuki,Mossel2010,Igloi-Rieger,Rigol2011,Calabrese-Essler,Essler-Caux,Brockmann,Pozsgay,Wouters,ProsenPRL2015,ProsenJune2015,Caux-Essler-Prosen}. 
Due to novel experimental demonstrations of out-of-equilibrium dynamics of  closed quantum systems made of cold atoms \cite{Kinoshita,prethermal,Fukuhara}, it has become an extensive research subject how thermalization occurs in isolated quantum systems 
\cite{Rigol-generic}. For integrable quantum systems,  which have a large number of conserved quantities, equilibration dynamics is not trivial. 
It has been conjectured that integrable quantum systems after long time the expectation values of local quantities approach those of the Generalized Gibbs Ensemble (GGE)  \cite{Rigol-GGE}. 
It has been investigated for  particular integrable models and initial states   
whether the GGE conjecture holds or not   \cite{Brockmann,Pozsgay,Wouters,ProsenPRL2015,ProsenJune2015,Caux-Essler-Prosen}.

The studies of non-equilibrium quantum dynamics are closely related to the recent renewed interest in the foundation of quantum statistical mechanics 
\cite{Tasaki98,Reimann08,Lebowitz2010,Reimann-Kastner,Lebowitz,Sugita,Reimann07,Sugiura}. 
One of the fundamental questions is 
whether and how equilibrium distribution functions are realized only through the unitary time evolution of a quantum state in an isolated system \cite{vonNeumann}. 
It has been demonstrated that in some general or concrete settings 
a pure initial state of an isolated quantum system 
equilibrates or thermalizes  in a certain mathematical sense 
\cite{Tasaki98,Reimann08,Lebowitz2010,Reimann-Kastner,Tasaki2013,TasakiNJP15,Jensen85,SKKD1}. 
It is  related to the idea that  a  typical pure state of a macroscopic quantum system can fully describe thermal equilibrium \cite{Lebowitz,Sugita,Reimann07,Sugiura}. The question  is often discussed in terms of the eigenstate thermalization hypothesis \cite{Deutsch,Srednicki}.

Although the above mentioned studies show that a wide class of isolated quantum systems 
thermalize or equilibrates to GGE, the time scale for thermalization has been studied only very recently \cite{Tasaki2013,TasakiNJP15}. Furthermore, the quantum dynamics of  a pair of spins  have been performed   numerically \cite{Evertz,Andrei} in an experimental setting \cite{Fukuhara}. However, it has not been explicitly shown how the expectation value of a certain local operator of an interacting integrable quantum system such as the Heisenberg chain of finite size equilibrates in time.  It has not been clear even whether it approaches some constant value or not.  Here we do not take the time average of the expectation value.  It is natural to conjecture that  as the system size increases,  the expectation value of a local operator at a given time becomes  close to the limiting value of the infinite system.  However, we expect that there are many aspects in the way how it approaches the equilibrium value in time and they are related to interesting topics of non-equilibrium dynamics close to integrable systems such as prethermalization  \cite{Berges} and pre-relaxation \cite{Bertini-Fagotti},  which should occur between the integrability and its breaking under non-integrable perturbations. Moreover, it is important to give a concrete example of equilibration of a local operator in a long period of time for interacting quantum systems from the viewpoint of  stability and equilibrium properties of KMS states \cite{Robinson}.

Let us introduce the Hamiltonian of the anti-ferromagnetic Heisenberg spin chain 
under the periodic boundary conditions (PBC):  $\sigma_{N+1}^{a} = \sigma_1^{a}$ 
for $a=X, Y, Z$.    
\be 
{\cal H} = {\frac 1 2} 
\sum_{j=1}^{N}\left\{ \sigma_j^{X} \sigma^{X}_{j+1} +  \sigma_j^{Y} \sigma^{Y}_{j+1} + 
 (\sigma_j^{Z} \sigma^{Z}_{j+1}  - 1) \right\} \, . 
\ee
Here, $\sigma_j^a$ denote the Pauli matrices. We also call the system the XXZ chain. 
In the present paper we show the exact relaxation dynamics of local magnetizations 
$\langle \sigma_m^z \rangle$ 
 of the spin-1/2 Heisenberg chain for an initially localized quantum state.   
We also study it for other states constructed from spinon states. 
Although it shows a strong oscillating behavior in time, 
we argue that the amplitudes of oscillation show a power-law decay,    
by evaluating the square deviations of the local magnetizations. 
We suggest the power-law decay may be universal 
for the expectation values of other local operators while the 
decaying exponent depends on the initial state.

The paper consists of the following. We briefly review the Bethe-ansatz equations   
and the scheme to evaluate local magnetizations,   
which is due to the recent development of the algebraic Bethe-ansatz  
\cite{Slavnov1989,Kitanine1999}. 
We first show that the relaxation time of the fidelity 
is given by the Boltmann time and is consistent with recent rigorous results.  
Here, the fidelity is  given by the square amplitude of correlation 
between the initial  state $|\Psi(0) \rangle$ and state at time $t$: 
$F(t)= |\langle \Psi(t)|  \Psi(0) \rangle |^2$. 
We then show the time evolution of the profile of local magnetizations $\langle \sigma_m^z \rangle$, and observe how initially localized profile collapses in time.  
It seems that the resulting oscillations of  local magnetizations $\langle \sigma_m^z \rangle$ continue forever. 
However, we shall show that the square deviations of the local magnetization exhibits  
a power-law decay. 

%
%
%
\section{Method} 
The eigenvectors and eigenvalues with $M$ down spins
of the Hamiltonian are constructed from solutions of  
the Bethe-ansatz equations (BAE) \cite{Korepin,TakaB}.  
\bea 
2 \tan^{-1} \left(  2 \lambda_{\alpha} \right) & = & {\frac {2 \pi} N}   J_{\alpha} 
+ {\frac 1 N} \sum_{\beta=1}^{M} 2 \tan^{-1} 
\left( \lambda_{\alpha} - \lambda_{\beta}   \right), \non \\ 
& &  \quad  \mbox{ for} \, \, \alpha= 1, 2 ,\ldots, M .  
\label{eq:BAElog}
\eea 
Here, quantum numbers $J_{\alpha}$ are given by integers or half-integers according to the condition:  $J_{\alpha} = (N-M+1)/2$  $ ({\rm mod}  \, 1)$. 
Here we take the branch: $|\tan^{-1} x| < \pi/2$ for any $x \in {\bm R}$.  
We denote by $| \{\lambda_j\} \rangle$ the eigenvector assoiated  with a solution of the BAE,  $\lambda_1, \lambda_2, \ldots, \lambda_M$. The energy eigenvalue $E$ of the eigenstate $| \{\lambda_j\} \rangle$ is  given by 
\be 
E_{\lambda} = -  \sum_{j=1}^{M} \frac 2 {1 + 4 \lambda^2_j} 
\, . 
\ee 

We consider a state $|\Phi \rangle$ consisting of a superposition of the Bethe-ansatz eigenvectors   $| \{\lambda_j\} \rangle$ 
\be 
| \Phi \rangle = \sum_{\lambda \in \Sigma} | \lambda \rangle c_{\lambda \Phi}  \, . 
\ee
Here we have abbreviated  eigenstate $| \{\lambda_j\} \rangle$ by 
$| \lambda \rangle$ and $\Sigma$ denotes a certain set of solutions of BAE.   

We denote by $\langle \sigma_m^z(t) \rangle $  the local magnetization on the $m$th site at time $t$, and define it by the expectation value of the local spin operator $\sigma_m^z$ 
with respect to the quantum state $| \Phi \rangle$ at time $t$  as follows. 
\be 
\langle \sigma_m^z(t) \rangle = \langle  \Phi (t) | \, \sigma^z_m \,  | \Phi(t) \rangle \, .  
\ee
Here the time evolution of the state $|\Phi(t) \rangle$ is given by 
$|\Phi(t) \rangle = \exp(-i H t/\hbar) |\Phi(0) \rangle$ where $|\Phi(0) \rangle$ 
is given by $| \Phi \rangle$. 
We evaluate the local magnetization by the form factor expansion:  
\bea 
& &  \langle  \Psi (t) | \, \sigma^z_m \,  | \Psi(t) \rangle  \non \\ 
& &  = \sum_{\lambda, \mu} 
\langle \mu|  \sigma_m^{z} | \lambda \rangle  \langle \Psi | \mu \rangle \langle \lambda | \Psi \rangle \exp \left( -(E_{\mu}- E_{\lambda})t  \right) \, .   \label{eq:FFE}
\eea
Here we recall that $E_{\lambda}$ denotes the energy  of the Bethe state $|\lambda \rangle$.  

In order to evaluate the matrix elements  $\langle \mu|  \sigma_m^{z} | \lambda \rangle$  
of the local operator $\sigma_m^z$ between two Bethe eigenstates $| \mu \rangle$ and 
$| \lambda \rangle$, we make use of the determinant formula of the matrix elements of  $\sigma_m^z$ derived by Kitanine, Maillet and Terras \cite{Kitanine1999}. For numerical evaluation, we further divide it out by the Cauchy determinant similarly as that of the 1D Bose gas \cite{Slavnov1989}.   

We remark that in order to show equilibration of local observables systematically, 
we may consider the Bloch vector, which is given by the expectation values of the traceless operators in the Hilbert space of the XXX spin chain with respect to a given quantum state \cite{Sugita}.  We consider the local magnetizations at sites $m$, $\langle \sigma_m^z \rangle$ for $m=1, 2, \ldots, N$ are  elements of the Bloch vector.   
%

\section{Fidelity} 

\begin{figure}[tb] 
\begin{center} 
\includegraphics[width=0.75\linewidth]{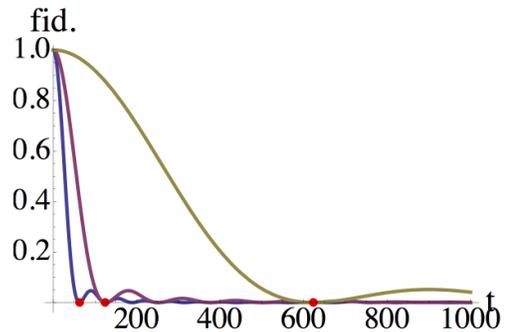}
 \end{center} 
\caption{
Time evolution of fidelity $F(t)$ for states $\Phi(0)\rangle $ given by sums of spinon states of equal weight in energy width $\Delta E$ for $\Delta E$ = 0.01, 0.05 and 0.1, respectively, in the XXX chain of $N=1, 000$ in the sector of $M=499$ down-spins.}   
\label{fig:fidelity}  
\end{figure}

Let us evaluate the fidelity for pure states $| \Phi(0) \rangle$ which are given the sums over spinon states with equal weight in the energy width $\Delta E$  for $\Delta = 0.01$, 0.05 and 0.1, etc.,  specifically. 
Each spinon eigenstate $| \lambda \rangle$ corresponds to a real solution of BAE (\ref{eq:BAElog}) specified  by putting  two holes 
in the set of  quantum integers $J_j$s \cite{Korepin,TakaB}.  
Here we assume that the system size is given by $N=1, 000$,  and the number of down-spins $M$ by half the system size minus 1: $M=499$.  
\be 
| \Phi(0) \rangle = \sum_{ \lambda \in S} c_{\lambda \Phi} | \lambda \rangle \,   
\ee
where symbol $S$ denotes a subset of the set of all spinon eigenstates, and 
in the case of equal weight we set  $c_{\lambda \Phi}= 1/\sqrt{|S|}$,  
where $|S|$ denotes the number of elements in the set $S$. 
We plot in Fig. \ref{fig:fidelity} the time evolution of the fidelity for the three states  
$| \Phi(t) \rangle$ after initial time $t=0$ with $\Delta E$  for $\Delta = 0.01$, 0.05 and 0.1:  $F(t)=  |\langle  \Phi(t) |\Phi(0) \rangle|^2 $.

We observe that the three graphs of fidelity $F(t)$ versus time $t$ shown 
in Fig. \ref{fig:fidelity}  are well approximated by the following expression \cite{Monnai}
\be 
F(t) = \frac 1 {t^2 + (\beta \hbar)^2}
\left( \frac {2 \hbar} {\Delta E} \right)^2 
\sin^2 \left( \frac {\Delta E} {2 \hbar} t \right) \, . 
\ee
There is only one fitting parameter $\Delta E$, practically. Here we assume that the inverse temperature is given by the energy width: $1/\beta = \Delta E$.  
The initial Gaussian behavior should be consistent with other approaches \cite{Prosen2002,Santos}. Non-exponential behavior in the fidelity of interacting many-body systems is recently argued \cite{delCampo}.

We estimate the relaxation time of fidelity through the numerical plots in Fig. \ref{fig:fidelity}.  In each graph we determine it by the time when the fidelity takes the minimum value first after initial time. The red dots in Fig. \ref{fig:fidelity} show the points of time when we determine the relaxation time $T_F$.

%
\begin{figure}[tb] 
\begin{center} 
\includegraphics[width=0.7\linewidth]{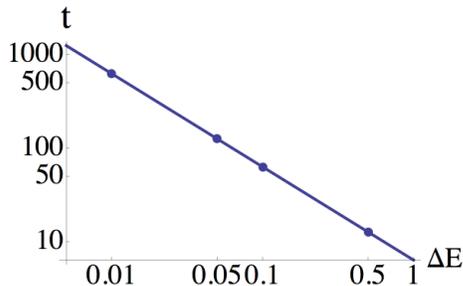}
 \end{center} 
\caption{
Relaxation time $t=T_F$ of fidelity for the states given by sums over spinon states with 
energy width $\Delta E$ 
and the line:  $T_F = h / \Delta E$. 
} 
\label{fig:Relaxation}  
\end{figure}

The estimates of relaxation time of fidelity $T_F$ are plotted in Fig. \ref{fig:Relaxation}. 
They are consistent with the line:  $T_F \approx h / \Delta E$. Here we remark that in the units of the present paper  we take $\hbar =1$. Thus,  it is given by the Boltzmann time 
assuming  $k_B T = \Delta E$.  Furthermore, it is interesting to observe that 
it coincides with a rigorous theoretical estimate  \cite{TasakiNJP15}  of thermalization time in a macroscopic quantum system for a typical non-equilibrium subspace: $T_R \approx h / \Delta E$.  

For the XXX chain of $N=10$ we evaluated the fidelity for considering 
all the Bethe-ansatz eigenstates associated with not only real solutions but also complex solutions as in Ref. \cite{Hagemans} (see also \cite{Giri-Deguchi2015}). 
We compared the time evolution of fidelity for 
the state consisting of only real solutions and that for the state of all solutions including complex solutions.  However, there is no clear difference in the time evolution of fidelity 
between the states of only real solutions and those of all solutions.


%
%
\section{Local magnetization} 

Let us now consider mainly the all-spinon state $|\Psi(0) \rangle$, which is given by the sum over all the spinon states for the XXX spin chain with an even system size $N$ in the sector of $M=N/2 -1$ down-spins, i.e. in the case of an almost half-filled lattice.  We evaluate the local magnetization $\langle \sigma_m^z(t) \rangle$ at time $t$ by eq. (\ref{eq:FFE}).

\begin{figure}[tb] 
\begin{flushright}
\includegraphics[width=0.95\linewidth]{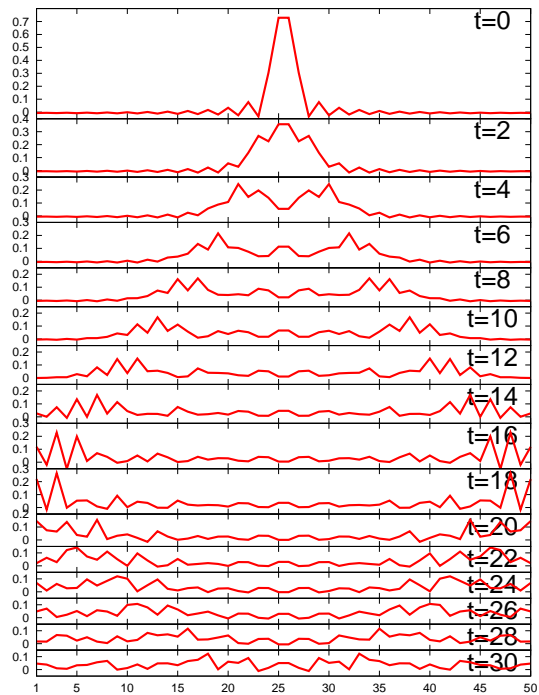}
\end{flushright}
\vskip 24pt
\caption{
Collapse of an initially  localized magnetization profile $\langle \sigma_m^z(t) \rangle$ 
($m=1, 2, \ldots, N$) in time evolution for the XXX chain with $N=50$ in the sector of $M=24$ down-spins. 
The profile is localized around at $m=25$ when $t=0$.  
} 
\label{fig:Collapse}  
\end{figure}

We observe in Fig. \ref{fig:Collapse}  that the  profile  of local magnetizations
 $\langle \sigma_m^z(0) \rangle$ for $m= 1, 2, \ldots, N$ at initial time $t=0$
is localized around at the middle site of $m=N/2$ and almost vanishes on other sites. 
We may consider it as  the  ``positive'' analogue of a ``quantum dark soliton'' constructed in the one-dimensional Bose gas with delta-function interactions \cite{SKKD1}, 
which are given by the sum over the type-II excitations in a branch.  
Here we remark that the average of $\sigma^z_m$ is given by 
$(N-2M)/N = 2/N = 0.04$ for $N=50$.

There are several procedures at different time steps as shown in Fig. \ref{fig:Collapse} 
for the whole time evolution in which the initially localized profile of local magnetizations 
$\langle \sigma_m^z(t) \rangle$ collapses to a uniformly oscillating profile in the final stage.   
First,  the initially-localized profile separates to two localized waves moving in different directions.  Then, the two localized waves propagate and collide each other at  $t=18 \sim 20$  around at the 1st site. They further propagate with smaller height of localized waves,    and finally they merge into such a profile that is roughly uniform in space 
but strongly oscillating in time.

\begin{figure}[tb] 
\begin{center} 
\includegraphics[width=0.9\linewidth]{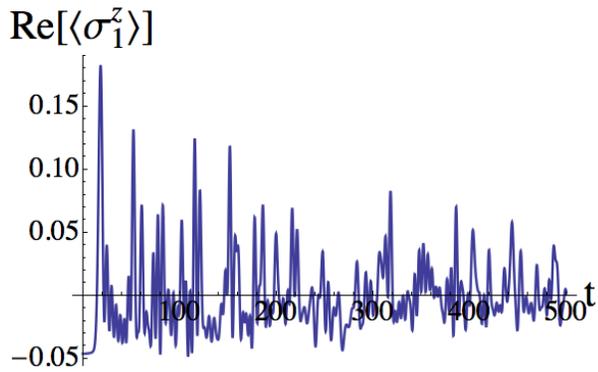}
 \end{center} 
\caption{
Slow relaxation of  local magnetization at site 1, $\langle \sigma_1^z(t) \rangle$,  
for the sum of all-spinon state in the XXX chain of $N=50$ with $M=24$.  
} 
\label{fig:fluctuation}  
\end{figure}

Thus,  the initially localized profile  in Fig. \ref{fig:Collapse} dynamically collapses to the roughly uniform profile oscillating in time.   
Here, the exact time evolution of the local magnetization at a site shows strong oscillating behavior in Fig. \ref{fig:fluctuation}.    
It seems that the fluctuations of the local magnetizations do not easily relax to the order of the inverse of the number of the Bethe-ansatz eigenstates consisting of the quantum state $|\Psi \rangle$, i.e.  $O(1/|S|)$, which is the order of magnitude expected statistically.

\begin{figure}[tb] 
\begin{center} 
\includegraphics[width=1.0\linewidth]{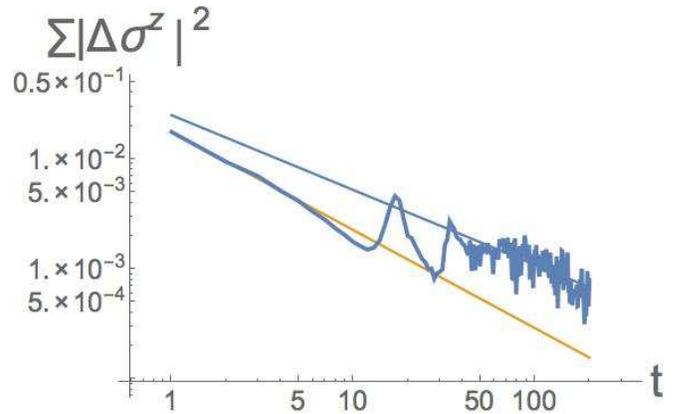}
 \end{center} 
\caption{
Relaxation of  the site-average of the squared deviations of the local magnetizations (\ref{eq:SD}) for $N=50$. Power law behavior such as $0.0250/t^{0.68}$ and 
$0.0176/t^{0.89}$ appears. } 
\label{fig:power-N50}  
\end{figure}

However,  by taking the site-average of the squared deviations of the local magnetizations 
over all sites we observe a power law decay over  a long period of time as in Fig.\ref{fig:power-N50}. 
We define the site-average of  squared deviations of local magnetizations 
on the $m$th sites $\langle \sigma_m^z \rangle$ by 
\be 
\langle  \left( \Delta \sigma^z(t) \right)^2 \rangle  
= {\frac 1 N} \sum_{m=1}^{N} \left( \langle \sigma_m^z(t) \rangle  
- {\frac 1 N}\sum_{j=1}^N  \langle \sigma_j^z(t) \rangle \right)^2 . 
\label{eq:SD}
\ee 
Here we remark that due to the symmetry of the Hamiltonian the sum of all the local magnetization is given by a constant: $\sum_{m=1}^{N} \sigma_m^z = N-2M$.

For the all-spinon state of $N=30$ , the squared deviations of local magnetizations decay almost as an inverse of time initially ($0.0289/t^{1.02}$), 
and as an inverse power of time with a smaller exponent ($0.0765/t^{0.806}$). 
For the all-spinon state of $N=50$  the squared deviations of local magnetizations 
are shown in Fig. \ref{fig:power-N50}.  For various other states which are  
given by the sums over sets of spinons, we observed similar power-law decay behavior of 
squared deviations (\ref{eq:SD}).    
Here, the estimates of the exponent are different among the states. For instance,  the yrast state, which is given by the sum over lowest excitations among spinon states, the estimate of the exponent is close to zero.  
We suggest that power-law decay may be universal in the time evolution
of the expectation values of local operators  for interacting integrable quantum systems, while the exponent depends on the states. 


%

We now argue that the time scale for equilibration of the local magnetizations is very long.  
We suggest that the time scale of the collapse of the initially localized  magnetization profile in the first stage is given by the system size divided by the spinon velocity. It is proportional to the system size. Here we recall that two localized waves travel in time in opposite directions in Fig. \ref{fig:Collapse}.  However,  the power-law decay of the squared deviations of the local magnetizations shown in Fig. \ref{fig:power-N50} 
shows  that there is no definite relaxation time in the whole long-time equilibration dynamics of the local magnetizations.

In summary we have performed exact relaxation dynamics of 
the local magnetization profile for the initially localized state $|\Psi \rangle$ in the XXX chain. Through the square deviations of local magnetizations 
we presented the  power-law decay of the local magnetization in a long time. 
We suggest that the expectation values of  other local operators may also decay
as a power of the time variable for interacting integrable quantum systems. 

\section*{Acknowledgement}
The authors would like to thank E. Kaminishi, S. Moriya and J. Sato for useful comments. 
The present study is partially supported by Grant-in-Aid for Scientific Research No. 15K05204.


\begin{thebibliography}{99}

\bibitem{McCoy1970} E. Barouch, B.M. McCoy and M. Dresden, 
Phys. Rev. {\bf 2}, 1075 
(1970).   




\bibitem{Rigol-GGE}
 M. Rigol, V. Dunjko, V. Yurovsky and M. Olshanii, 
 Phys. Rev. Lett {\bf 98}, 050405 (2007);

\bibitem{Rigol-generic}
 M. Rigol, V. Dunjko and M. Olshanii, Nature {\bf 452}, 854 (2008).



\bibitem{SeiSuzuki} D. Rossini, S. Suzuki, G. Mussardo, G.E. Santoro and 
A. Silva, Phys. Rev. B {\bf 82}, 144302 (2010).    

\bibitem{Mossel2010} J. Mossel and J.-S. Caux, 
New J. Phys. {\bf 12}, 055028 (2010).  


\bibitem{Igloi-Rieger} F. Igl{\'o}i and H. Rieger, Phys. Rev. Lett. {\bf 106}, 035701 (2011).   

\bibitem{Rigol2011} A.C. Cassidy, C.W. Clark, and M. Rigol, 
Phys. Rev. Lett. {\bf 106}, 140405 (2011).  


\bibitem{Calabrese-Essler} P. Calabrese, F.H.L. Essler and M. Fagotti, 
Phys. Rev. Lett. {\bf 106}, 227203 (2011); J. Stat. Mech. (2012) P07016; 
J. Stat. Mech. (2012) P07022. 

%
\bibitem{Essler-Caux} J.-S. Caux and F.H.L. Essler, Phys. Rev. Lett. {\bf 110}, 257203 (2013).  

\bibitem{Brockmann} M. Brockmann, B. Wouters, D. Fioretto, J. De Nardis, R. Vlijm and J.-S. Caux,  J. Stat. Mech., P12009 (2014).   
%

\bibitem{Pozsgay} B. Pozsgay, M. Mesty{\'a}n, M.A. Werner, M. Kormos, G. Zar{\'a}nd and G. Tak{\'a}cs, 
Phys. Rev. Lett. {\bf 113}, 117203 (2014).   
%


\bibitem{Wouters} B. Wouters,  J. De Nardis, M. Brockmann, D. Fioretto, M. Rigol, and J.-S. Caux, Phys. Rev. Lett. {\bf 113}, 117202 (2014).   

\bibitem{ProsenPRL2015} M. Mierzejewski, P. Prelo{\v v}sek and T. Prozen, Phys. Rev. Lett. 
{\bf 114}, 140601 (2015).  

\bibitem{ProsenJune2015} E. Ilievski, M. Medenjak and T. Prosen, 
arXiv:1506.05049 .  

\bibitem{Caux-Essler-Prosen}  
E. Ilievski, J. De Nardis, B. Wouters, J.-S. Caux, F.H.L. Essler, and T. Prosen, 
arXiv:1507.02993 . 




\bibitem{Kinoshita} T. Kinoshita, T. Wenger and D.S. Weiss, 
Science {\bf 305}, 1125 (2004); Phys. Rev. Lett. {\bf 95}, 190406 (2005); 
Nature {\bf 440}, 900 (2006).  

\bibitem{prethermal} M. Gring, M. Kuhnert, T. Langen, T. Kitagawa, B. Rauer, M. Schreitl, I. Mazets, D. Adu Smith, E. Demler, J. Schmiedmayer, 
Science {\bf 337}, 1318 (2012).  

\bibitem{Fukuhara} T. Fukuhara, P. Schau\ss, M. Endres, S. Hild, 
M. Cheneau, I. Bloch and C. Gross, Nature {\bf 502}, 76 (2013).  

%
%

\bibitem{Tasaki98}
H. Tasaki, Phys. Rev. Lett. {\bf 80}, 1373 (1998).

\bibitem{Lebowitz2010}
S. Goldstein, J.L. Lebowitz, 
C. Mastrodonato, R. Tumulka, and N. Zanghi,  Phys. Rev. E {\bf 81}, 011109 (2010). 


\bibitem{Reimann-Kastner} P. Reimann and M. Kastner, New J. Phys. {\bf 14}, 043020 (2012) 



%
\bibitem{Lebowitz}
S. Goldstein, J.L. Lebowitz, R. Tumulka, and N. Zanghi,  Phys. Rev. Lett. {\bf 96}, 050403 (2006)

\bibitem{Sugita}
A. Sugita, Nonlinear Phenom. Complex Syst. {\bf 10}, 192 (2007).


\bibitem{Reimann07} 
P. Reimann, Phys. Rev. Lett. {\bf 99}, 160404 (2007).

\bibitem{Reimann08} 
P. Reimann, Phys. Rev. Lett. {\bf 101}, 190403 (2008).


\bibitem{Sugiura} S. Sugiura and A. Shimizu, Phys. Rev. Lett. {\bf 108}, 240401 (2012);  
Phys. Rev. Lett. {\bf 111}, 010401 (2013).  






\bibitem{vonNeumann}
R. Tumulka, Eur. Phys. J. H {\bf 35}, 201 (2010)  
[J. von Neumann, Z. Phys. 57, 30 (1929)]



\bibitem{Tasaki2013} S. Goldstein, T. Hara and H. Tasaki, Phys. Rev. Lett. {\bf 111}, 140401 (2013).  

\bibitem{TasakiNJP15} S. Goldstein, T. Hara and H. Tasaki, New J. Phys. {\bf 17}, 045002 (2015) 



\bibitem{Jensen85} R.V. Jensen and R. Shanker,  , Phys. Rev. Lett. {\bf 54}, 1879 (1985). 


\bibitem{SKKD1} J. Sato, R. Kanamoto, E. Kaminishi, and T. Deguchi, 
Phys. Rev. Lett. {\bf 108}, 110401 (2012). 







%
%
\bibitem{Deutsch} J. M. Deutsch, Phys. Rev. A {\bf 43}, 2046 (1991). 
\bibitem{Srednicki} M. Srednicki, Phys. Rev. E {\bf 50}, 888 (1994). 






%
%
\bibitem{Evertz} M. Ganahl, E. Rabel, F. H. L. Essler, and H. G. Evertz, Phys. Rev. Lett. {\bf 108}, 077206 (2012).   
\bibitem{Andrei} W. Liu and N. Andrei, Phys. Rev. Lett. {\bf 112}, 257204 (2014).  



%
%
\bibitem{Berges} J. Berges, S. Borsanyi, C. Wetterrich, Phys Rev. Lett. {\bf 93}, 
 142002 (2004)

\bibitem{Bertini-Fagotti} B. Bertini and M. Fagotti, arXiv:1501.07260 
%
%



%
%

\bibitem{Robinson} O. Bratteli and D.W. Robinson, {\it Operator Algebras and Quantum Statistical Mechanics 2: Equilibrium States} (Berlin Heidelberg New York: Springer-Verlag)  
(1997). 

%
%
\bibitem{Kitanine1999}   
N. Kitanine, J.M. Maillet and V. Terras, 
Nucl. Phys. B {\bf 554} [FS] (1999) 647--678 

\bibitem{Slavnov1989} N. A. Slavnov, Teor. Mat. Fiz. {\bf 79}, 232 (1989) 
; {\bf 82}, 389 (1990); 




%
%

\bibitem{Korepin}  V. Korepin, N. Bogoliubov and A. Izergin, {\it ``Quantum Inverse Scattering Method and Correlation Functions"} 
(Cambridge: Cambridge University Press)  (1993).

\bibitem{TakaB}  M. Takahashi,  {\it ``Thermodynamics of One-Dimensional Solvable Models"},  (Cambridge: Cambridge University Press) (1999).

%
%


\bibitem{Monnai} T. Monnai, J. Phys. Soc. Jpn. {\bf 83}, 064001 (2014). 

\bibitem{Prosen2002} T. Prosen, Phys. Rev. E {\bf 65}, 036208 (2002).   

\bibitem{Santos} E. J. Torres-Herrera and Lea F. Santos, Phys. Rev. A {\bf 90}, 033623 (2014)  
\bibitem{delCampo} A. del Campo, arXiv:1504.01620 .

%
%

\bibitem{Hagemans} R. Hagemans and J.-S. Caux, J. Phys. A: Math. Theor.  {\bf 40}  14605-14647 (2007).

\bibitem{Giri-Deguchi2015}   
P. R. Giri and T. Deguchi,   
J. Stat. Mech., P07007 (2015) 




\end{thebibliography}
\end{document}